# Harnessing hidden quantum metric response in a 2D magnet via nonlocal photovoltaic effect

Yong Tan[1,2#], Qian Hu[1,3#], Rui-Chun Xiao[4,5#], Hang Zhou[6], Yuqing Huang[1,3*], Zelalem Abebe Bekele[1], Yongcheng Deng[1] , Xuan Qian[1,3], Qikang Gan[7], Lei Wang[7], Yang Ji[8], Ding-Fu Shao[6*], Lixia Zhao[2*], Kaiyou Wang[1,3*]

[1] State Key Laboratory of Semiconductor Physics and Chip Technologies, Institute of Semiconductors, Chinese Academy of Sciences, Beijing, China

[2] School of Electronics and Information Engineering, Tiangong University, Tianjin, China

[3] College of Materials Science and Opto-Electronic Technology, University of Chinese Academy of Sciences, Beijing, China

[4] Institute of Physical Science and Information Technology, Anhui University, Hefei 230601, China

[5] Anhui Provincial Key Laboratory of Magnetic Functional Materials and Devices, Anhui University, Hefei 230601, China

[6] Key Laboratory of Materials Physics, Institute of Solid State Physics, Hefei Institutes of Physical Science, Chinese Academy of Sciences, Hefei 230031, China

[7] National Laboratory of Solid-State Microstructures, School of Physics and Collaborative Innovation Center of Advanced Microstructures, Nanjing University, Nanjing, China

[8] School of Physics, Zhejiang University, Hangzhou, China

[#]These authors contribute equally.

*Corresponding author. Email: yqhuang@semi.ac.cn (Yuqing Huang), dfshao@issp.ac.cn (Ding-Fu Shao), lxzhao@tiangong.edu.cn (Lixia Zhao), kywang@semi.ac.cn (Kaiyou Wang)

**The quantum geometry of Bloch wavefunctions underpins a wealth of emergent phenomena in quantum materials. Its imaginary part, the Berry curvature, has long been recognized as a key source for hallmark effects such as quantum Hall and topological phenomena, etc. The real part of quantum geometry, the quantum metric, has recently garnered considerable attention due to predictions of a range of unconventional nonlinear and nonequilibrium responses. Such responses usually vanish in centrosymmetric systems, largely restricting relevant studies to non-centrosymmetric materials. Here we challenge this convention by revealing that the vanished quantum metric response can survive in a hidden form. Using a non-local photovoltaic scheme in a layered magnetic semiconductor, we spatially separate mutually compensating photocurrents and thereby detect such hidden quantum metric response. We demonstrate this effect across distinct magnetic states and down to the ultrathin limit. Moreover, we realize reconfigurable, nonvolatile and probabilistic photodetection enabled by the quantum metric response. These results not only fundamentally expand the material landscape for quantum geometric physics, but also open new gateway to harvest the quantum geometric contributions for state-of-the-art nonvolatile reprogrammable sensing and computing applications.**

In condensed matter physics, the geometry of Bloch wavefunctions in momentum space underpins a wealth of emergent phenomena. The imaginary part of the quantum geometric tensor, the Berry curvature, has been extensively studied for its pivotal role in generating topological effects such as integer, fractional and anomalous quantum Hall effects, chiral anomalies, etc. [1-8]. On the contrary, the real part of the quantum geometric tensor, the so-called quantum metric, has remained far less explored until very recently, when theoretical advances have emerged and predicted a wealth of exotic quantum-metric-driven nonlinear and nonequilibrium responses, including the nonlinear Hall effect (NHE) [9-15], bulk photovoltaic effect (BPVE) [16-22], etc. Furthermore, the effect of quantum metric on superconductivity, photocurrent and anomalous magnetic response in flatband systems has also been addressed lately [23-26].

A key constraint for the macroscopic observation of quantum-metric-governed effects is crystal symmetry. Specifically, a non-vanishing quantum metric response, such as a DC photocurrent from BPVE, typically requires the breaking of inversion symmetry ($\mathcal{P}$). This requirement has, to date, largely confined the exploration and application of quantum-metric-related effects to noncentrosymmetric materials, or to systems with $\mathcal{P}$-symmetry deliberately broken via external fields [11-14,27]. This principle is elegantly demonstrated in A-type antiferromagnets [28]. For instance, in even-layered structures where both $\mathcal{P}$ and time-reversal symmetry ($\mathcal{T}$) are individually broken but the combined $\mathcal{PT}$ operation is preserved, a finite quantum metric can emerge, enabling the aforementioned nonlinear phenomena [11-14,17-21].

Conversely, in centrosymmetric magnetic phases, e.g. ferromagnets or odd-layered A-type antiferromagnets where $\mathcal{P}$ is globally preserved, the prevailing symmetry analysis predicts a vanishing even-order quantum metric response including BPVE and NHE. This has led to the general perception that such centrosymmetric magnets are irrelevant to quantum metric physics.

In this work, we overturn this conventional understanding by demonstrating that a substantial quantum metric response survives in a "hidden" form in centrosymmetric 2D magnets. While the global inversion symmetry strictly forbids its macroscopic manifestation, we reveal that these latent geometric properties can be activated, probed, and harvested through the photovoltaic effect. By exploiting a non-equilibrium and non-local photocurrent detection scheme, we dynamically unlock a pronounced photovoltaic response and showcase its potential for encoding nonvolatile and reprogrammable intelligent photodetection. These findings fundamentally extend the material boundary for quantum-metric-driven phenomena, unveiling a vast landscape of hidden physics within symmetric phases and enabling new design principles for optoelectronic and spintronic devices.

**A non-equilibrium, non-local access to hidden quantum metric response**

The quantum metric $g$ is a fundamental geometric property of Bloch electrons that characterizes the distance between neighboring quantum states in momentum space. To manifest this intrinsic geometric attribute through macroscopic transport, electrons carrying finite quantum metric must acquire a group velocity $\boldsymbol{v}$ along the transport direction. The interplay between the carrier's velocity and its inherent geometry gives rise to the characteristic quantity $\boldsymbol{v} \cdot g$, whose Brillouin-zone integration defines the quantum metric dipole (QMD) and is regarded as the fundamental source driving second-order nonlinear responses.

Since $g$ and $\boldsymbol{v}$ are respectively even and odd functions under both $\mathcal{T}$ and $\mathcal{P}$, the QMD is inherently $\mathcal{T}$- and $\mathcal{P}$-odd. Therefore, emergence of a finite QMD and hence the net macroscopic response can only exist in noncentrosymmetric magnetic systems where both $\mathcal{T}$ and $\mathcal{P}$ are broken. For example, as illustrated in Fig. 1a for a centrosymmetric magnet, the $\mathcal{P}$-symmetry forces $\boldsymbol{v} \cdot g$ to distribute perfectly antisymmetrically in momentum space. Consequently, the current $\boldsymbol{J}$ arising from QMD, whether driven by an electric field or generated by optical excitation, consists of two components (denoted as $\boldsymbol{J}_1$ and $\boldsymbol{J}_2$ for simplification) that are equal and opposite ($\boldsymbol{J}_1 = -\boldsymbol{J}_2$), thereby canceling exactly and yielding a vanishing net macroscopic current. By contrast, in Fig. 1b when inversion symmetry is broken, $\boldsymbol{v} \cdot g$ distribution becomes imbalanced in momentum space, resulting in $\boldsymbol{J}_1 \neq -\boldsymbol{J}_2$ and a finite net current.

Crucially, the two mutually-compensating currents $\boldsymbol{J}_1$ and $\boldsymbol{J}_2$ hidden in $\mathcal{P}$-preserved systems are intimately connected by global $\mathcal{P}$-symmetry. A key observation is made that $\boldsymbol{J}_1$ and $\boldsymbol{J}_2$ in real space predominantly reside in two $\mathcal{P}$-related sectors of the material. Therefore, if a spatial decoupling scheme can be established that allows to distinguish and independently address these sectors, it becomes conceptually feasible to selectively harvest the hidden current components.

Achieving such scheme depends critically on the chosen transport regime, which is markedly different for the two most prominent quantum-metric-governed phenomena, namely NHE and BPVE. While NHE serves as a powerful probe of quantum geometry, it typically manifests in metallic systems. Large conductivity makes it virtually impossible to electrically isolate the $\mathcal{P}$-related sectors in metallic samples. By contrast, the BPVE operates efficiently in semiconductors or insulators. Under optical excitation, $\boldsymbol{J}_1$ and $\boldsymbol{J}_2$ emerge as non-equilibrium hot-carrier photocurrents while the bulk of the material maintains its intrinsic insulating character. It prevents internal short-circuiting of the two hidden currents, spatially confining $\boldsymbol{J}_1$ and $\boldsymbol{J}_2$ to their respective $\mathcal{P}$-related sectors. Designing a spatially confined, non-local BPVE detection scheme therefore enables exclusively harvesting either $\boldsymbol{J}_1$ or $\boldsymbol{J}_2$.

Specifically, we consider an unconventional detection scheme for the BPVE injection current in a layered magnetic material as illustrated in Fig. 1c, where electrodes contact solely the bottom surface of the material. The optical excitation is confined to sample regions away from the electrodes, thereby preserving the global $\mathcal{P}$-symmetry of the photocurrent generation process. It generates equal-magnitude counter-propagating hidden photocurrents $\boldsymbol{J_1}$ and $\boldsymbol{J_2}$, which are detected without any external electrical bias. These electrodes predominantly collect the photocurrent originating from the bottom layers (e.g., $J_1$). This non-equilibrium, non-local photovoltaic scheme thus unlocks the hidden quantum metric response that would otherwise remain perfectly concealed in centrosymmetric magnetic materials.

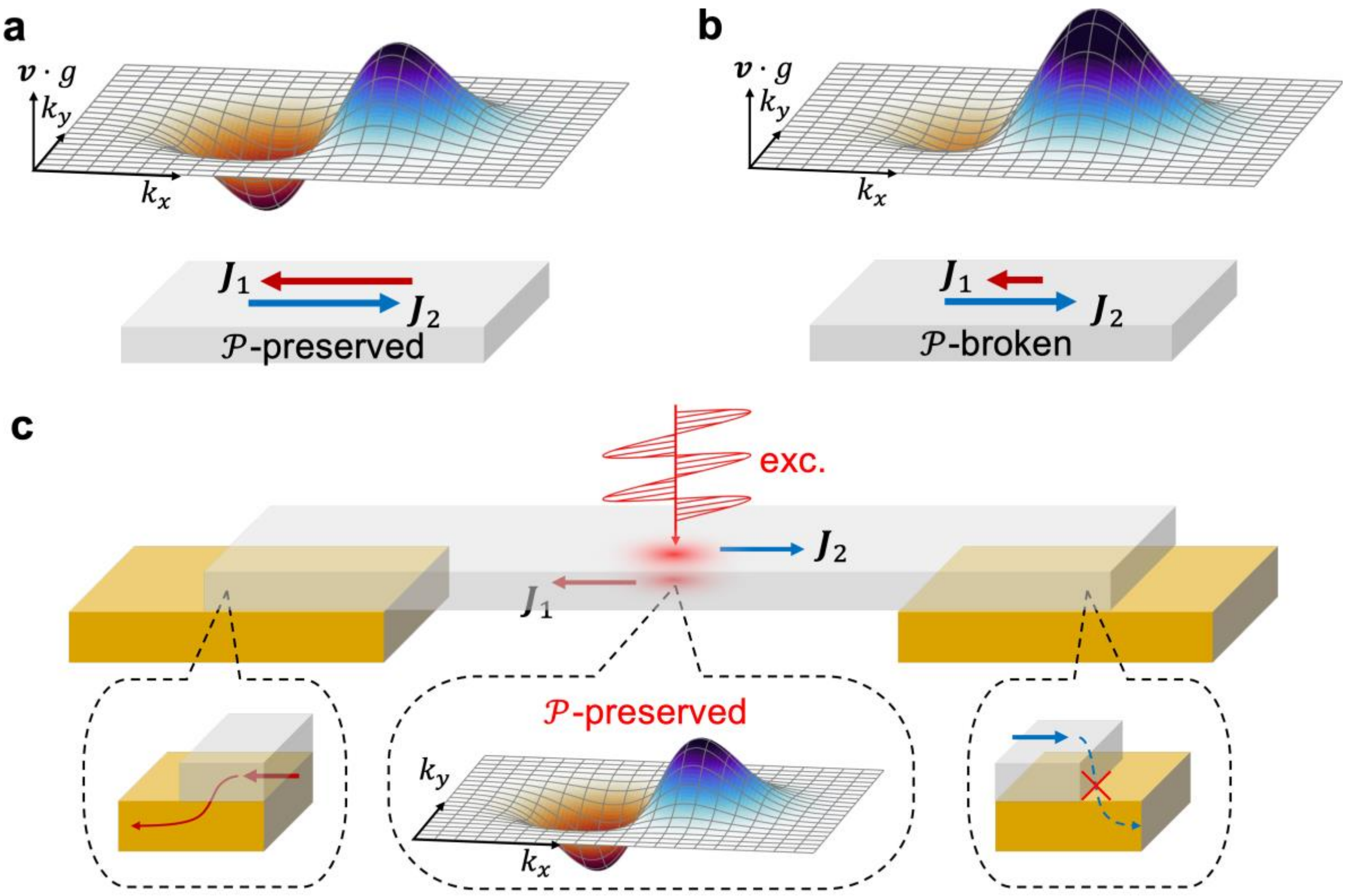

**Fig. 1. Hidden quantum metric response probed by photocurrent.** **a**, Illustration of vanished quantum metric current response in $\mathcal{P}$-preserved system. The k-space integration quantum metric dipole (QMD) $\boldsymbol{v}.g$ vanishes. $\boldsymbol{J}_1$ and $\boldsymbol{J}_2$ are the current response carrying positive and negative QMD. **b**, Illustration of finite quantum metric current response in $\mathcal{P}$-broken system. **c**, Experimental scheme for probing the hidden quantum metric response in $\mathcal{P}$-preserved materials.

**Observation of the hidden quantum metric response in a layered magnet**

Following the proposed scheme, we select CrSBr as a model system to demonstrate detection of the hidden quantum metric response. CrSBr is an intrinsic 2D magnetic semiconductor which has attracted intense attention due to its unique properties, including strong light-matter interaction, magneto-electronic coupling, nonlinear magnons, etc. [29-34]. It has a centrosymmetric lattice structure (Fig. 2a) and an A-type antiferromagnetic (AFM) ground state with the Cr spin in the neighboring layers aligned anti-parallel along the crystallographic $b$-axis (Fig. 2b). The AFM spin configuration can be converted to ferromagnetic (FM) by applying magnetic fields along $b$-axis, which is the magnetic easy axis. For odd-layered CrSBr, the $\mathcal{T}$-symmetry is broken while the $\mathcal{P}$-symmetry is preserved for both AFM and FM orderings. This leads to a globally-vanished QMD, making it an ideal platform for exploring the hidden quantum metric response.

To begin with, the photovoltage response of a 5L-CrSBr device (Fig. 2c) is characterized using lock-in technique (see method and Extended Data Fig. 1). Fig. 2d shows the reflectance spectra for various magnetic fields with the incident light polarization parallel to $b$-axis. The magnetoreflectance spectra reveal two excitonic states, denoted as $X_1$ and $X_2$, which redshift when switching from AFM to FM and are attributed to bulk exciton and magnetically confined surface exciton in earlier reports [35, 36]. Fig. 2e and 2f show the photovoltage spectra detected along $a$-axis electrodes under the same excitation condition as in the magnetoreflectance measurements. The photovoltage is found controllable with magnetic fields and the contributions of $X_1$ and $X_2$ excitons are evident in the spectra.

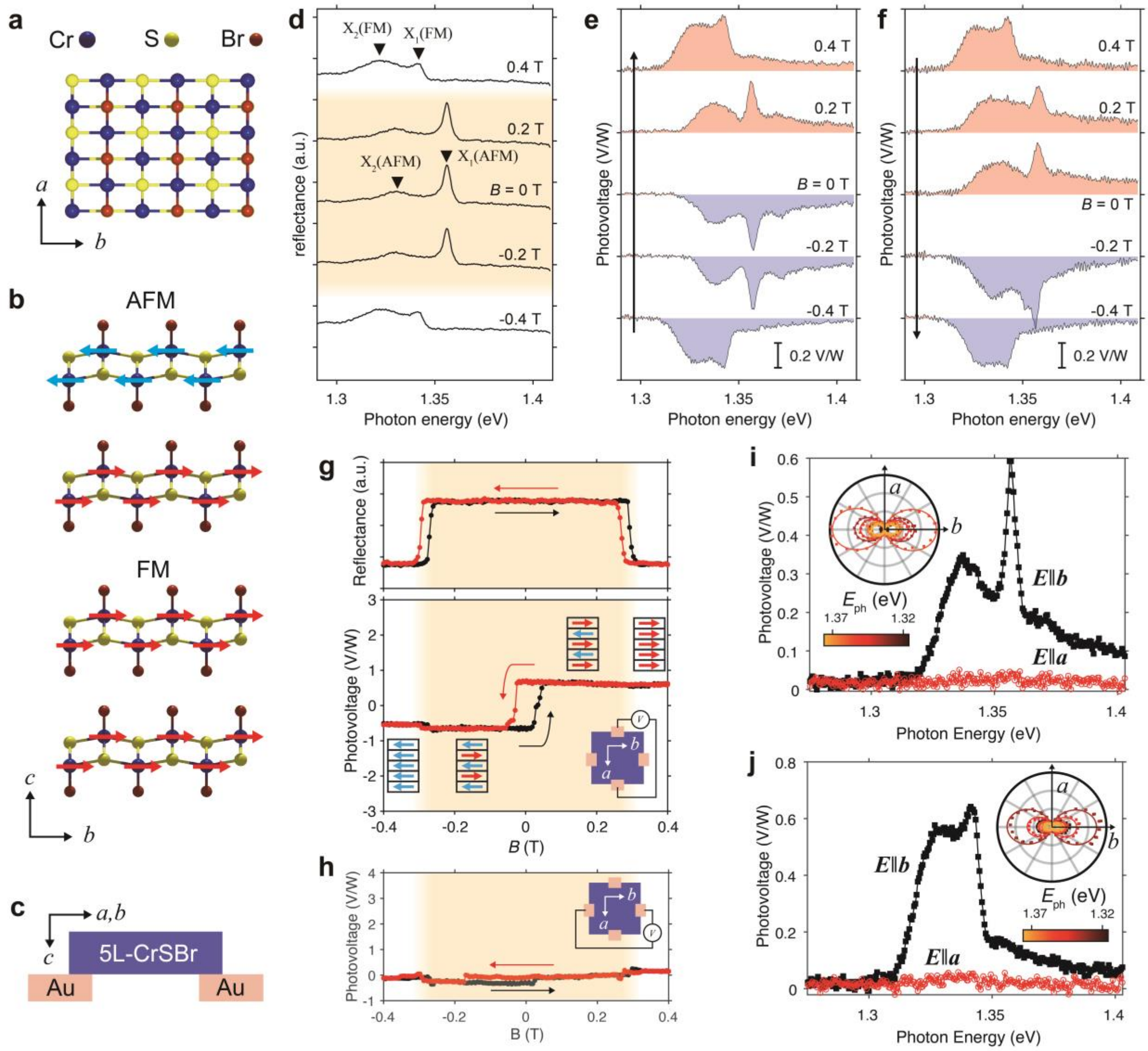

**Fig. 2. Detection of the hidden quantum metric response in a 5L-CrSBr.** **a**, Top view lattice structure of CrSBr. **b**, Side view lattice structure of CrSBr and the corresponding spin configurations for AFM and FM states. **c**, Schematic photovoltage measurement scheme for 5L-CrSBr device. **d**, Magnetoreflectance spectra reveal two dominant excitonic resonances $X_1$ and $X_2$, which shift in energy between AFM and FM states. **e** and **f** are the photovoltage spectra under forward and backward magnetic field sweep. **g**, Magnetoreflectance (upper panel) and photovoltage detected along $a$-axis (lower panel) as function of *B*. The plausible magnetic states are marked and measurement scheme is shown in the inset. **h**, Photovoltage detected along $b$-axis as function of $B$ with the corresponding measurement scheme shown in the inset.

**i** and **j** are the photovoltage spectra under different excitation light polarization for AFM ($B$ = 0 T) and FM ordering ($B$ = 0.4 T). Inset: Polarization angular dependence of the photovoltage for selected excitation photon energies $E_{\mathrm{ph}}$. All the measurements are performed at 10 K.

Next, we verify that the magnetically dependent photovoltages arise from hidden QMD in CrSBr. To begin with, as the QMD is a $\mathcal{T}$-odd function, the photovoltaic response driven by QMD must reverse its sign upon spin flips. This is exactly found in Fig. 2e and 2f, where a positive (negative) photovoltage is detected at positive (negative) magnetic fields. Additionally, a non-volatile behavior is observed that the photovoltage persists after removal of the external magnetic field and the zero-field photovoltage depends on the field-sweeping history. In the lower panel of Fig. 2g, we plot the photovoltage as a function of magnetic and compare with the magnetoreflectance result in the upper panel. The photovoltage sums up the obtained values under resonant excitation of $X_1$(FM) and $X_1$(AFM) with excitation photon energies of 1.342 eV and 1.356 eV. The plausible magnetic states are sketched in the insets of Fig. 2g. It demonstrates that the polarity of photovoltage is reversed between spin-flipped magnetic states for both FM and AFM orderings. Secondly, the photovoltage correlates strongly with the spin configuration of the bottom surface monolayer and reverses its sign upon flipping of the surface-layer magnetization at $\pm 0.03$ T as triggered by the Zeeman energy of the uncompensated magnetization in an odd-layer CrSBr [37]. This critical behavior is also consistent with the hidden quantum metric mechanism [22].

For a symmetric perspective, the observed photovoltage arise from QMD component $v_x g_{yy}$, which is the dominant QMD component as revealed by the first-principles calculations (see method and Extended Data Fig. 2). By contrast, measurement along $b$-axis yields a negligible signal as $v_y g_{yy}$ is forbidden by symmetry (Fig. 2h). The magnetic photovoltage also vanishes when the polarization of the incident light aligns with the $a$-axis (Fig. 2i and 2j), which is forbidden both by band-edge transition selection rules of CrSBr [38,39] as well as the proposed hidden quantum metric mechanism.

**Spatial mapping and layer dependence of the quantum metric response**

Our non-local approach facilitates the spatial mapping of the quantum metric response. This is achieved by scanning the excitation spot while monitoring the photovoltage in the 5L-CrSBr device (Fig. 3a). Fig. 3b presents the measurement results under different magnetic fields, along with the corresponding magnetic field-sweeping history. The photovoltage obtained at FM state ($B$ = + 0.4 T and - 0.4 T) shows opposite polarity for spin-up ($|\bar{\uparrow}\uparrow\uparrow\uparrow\underline{\uparrow}\rangle$) and spin-down state ($|\bar{\downarrow}\downarrow\downarrow\downarrow\underline{\downarrow}\rangle$) and is generated uniformly within the whole sample area. Here, $\bar{\uparrow}$ ($\bar{\downarrow}$) and $\underline{\uparrow}$($\underline{\downarrow}$) denote the up- (down-) spin state along $b$-axis for top and bottom surface monolayer. The magnetic photovoltage mapping at $B$ = 0 T for Néel spin-up and -down states (e.g. $|\bar{\uparrow}\downarrow\uparrow\downarrow\underline{\uparrow}\rangle$ and $|\bar{\downarrow}\uparrow\downarrow\uparrow\underline{\downarrow}\rangle$) also shows uniformly distributed and $\mathcal{T}$-odd signal after subtracting a spin-state-independent photovoltage component (see Extended Data Fig. 3) due to the CrSBr/Au Schottky junction (see method and Extended Data Fig. 4-5 for details). Similar results are also found in a 4L-CrSBr device (see Extended Data Fig. 6), albeit the fact that the even-layered CrSBr has a broken $\mathcal{P}$-symmetry in its AFM ground states that enables the conventional quantum metric response. The magnetic photovoltage and hence the quantum metric response in both 5L- and 4L-CrSBr devices distribute uniformly throughout the sample area for both FM and AFM states and even with disparate global symmetries. These results corroborate the common origin between the conventional and hidden quantum metric response, providing unambiguous evidence for experimental observation of the intrinsic response originating from the hidden quantum metric.

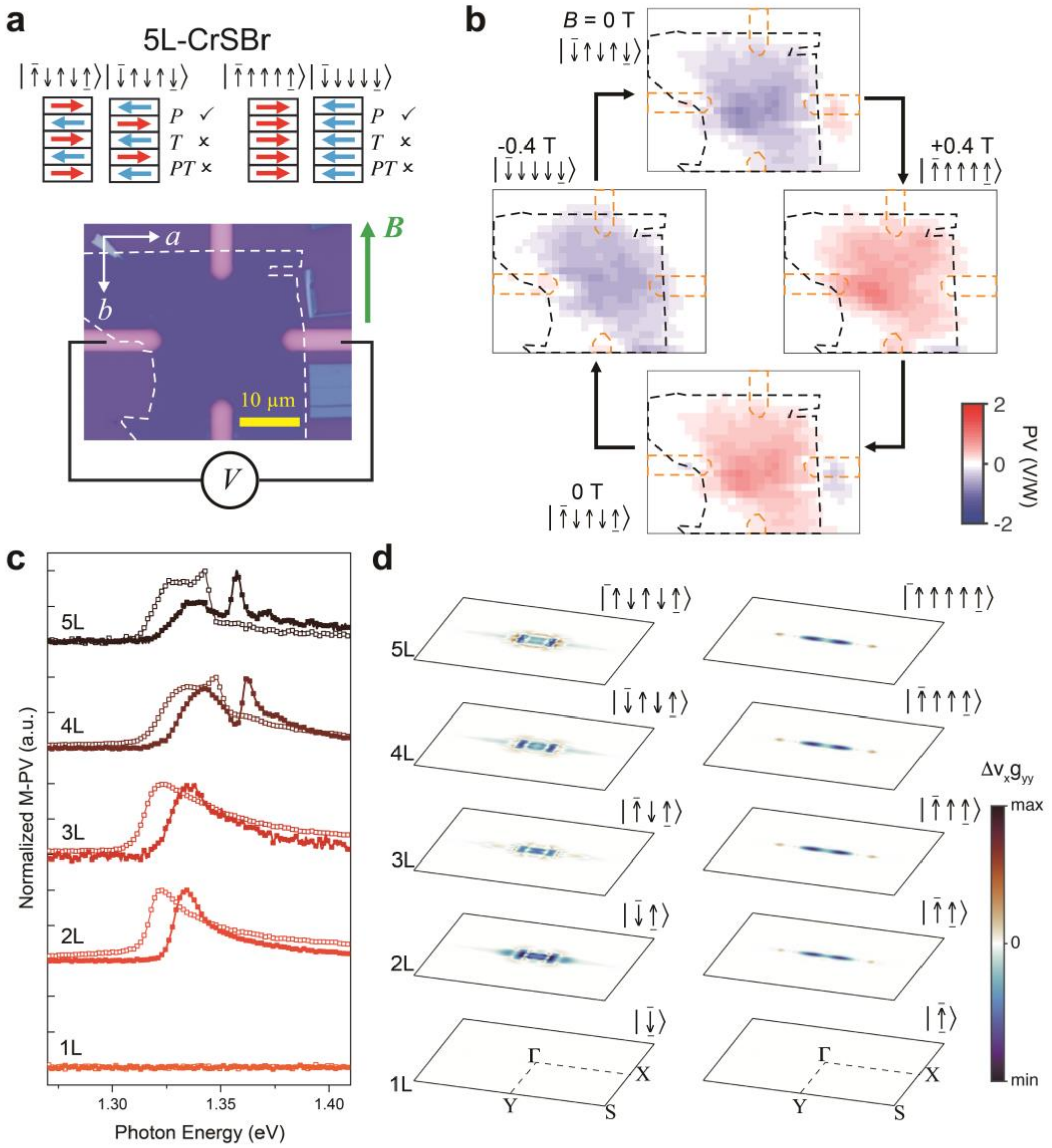

**Fig. 3. Spatial Mapping and layer dependence of the magnetic photovoltaic response. a,** Optical microscopy image of the 5L-CrSBr device. The plausible AFM and FM states are illustrated with the corresponding symmetries. **b**, Spatial mapping of the magnetic photovoltage in 5L-CrSBr under resonant excitation of $X_1$ and its evolution with magnetic fields. **c**, The normalized magnetic photovoltage (M-PV) spectra for FM (open symbols) and AFM (filled symbols) for CrSBr devices with different layers. **d**, The calculation of the momentum-space distribution of the hidden QMD $\Delta v_x g_{yy}$ in bottom surface monolayer for CrSBr slab with different layer numbers and spin states.

Fig. 3c displays the magnetic photovoltage spectra of devices with different layers, where the results for the FM and AFM states are plotted with open and filled symbols, respectively. Evidently, the magnetic photovoltage is consistently observed across devices of different layer numbers, down to the bilayer sample. The conventional BPVE should vanish in all cases except for AFM state of even-layered CrSBr as shown by the perfectly antisymmetrically-distributed QMD $v_x g_{yy}(\mathbf{k})$ in Extended Data Fig. 7. Nevertheless, the hidden QMD featured by an imbalanced momentum distribution of $v_x g_{yy}(\mathbf{k})$ in $\mathcal{P}$-related sectors universally exists in such centrosymmetric magnetic phases as shown by the calculation results in Extended Data Fig. 8. We note that the magnetic photovoltage spectra in Fig. 3c are obtained by differentiating the photovoltage detected for opposite spin states, effectively removing any nonmagnetic contributions. Following a similar principle, Fig. 3d displays the calculated momentum-dependent hidden QMD, denoted as $\Delta v_x g_{yy}(\mathbf{k}) = v_x g_{yy}(\mathbf{k}) - \mathcal{T}(v_x g_{yy}(\mathbf{k}))$, for the bottom surface monolayer of CrSBr slabs across various layer numbers and spin states. $\Delta v_x g_{yy}(\mathbf{k})$ therefore highlights the hidden QMD component responsible for the magnetic photovoltage. The fact that the magnetic photovoltage in accordance with the calculated $\Delta v_x g_{yy}(\mathbf{k})$ is detected in all devices except for the monolayer sample, where the layer degree of freedom is no longer applicable and the $\mathcal{P}$-related sectors are strongly correlated, further highlights the capability of our approach to probe the hidden quantum metric effect down to extreme thickness in a layered 2D magnet.

**Nonvolatile and reprogrammable photodetector driven by quantum geometry**

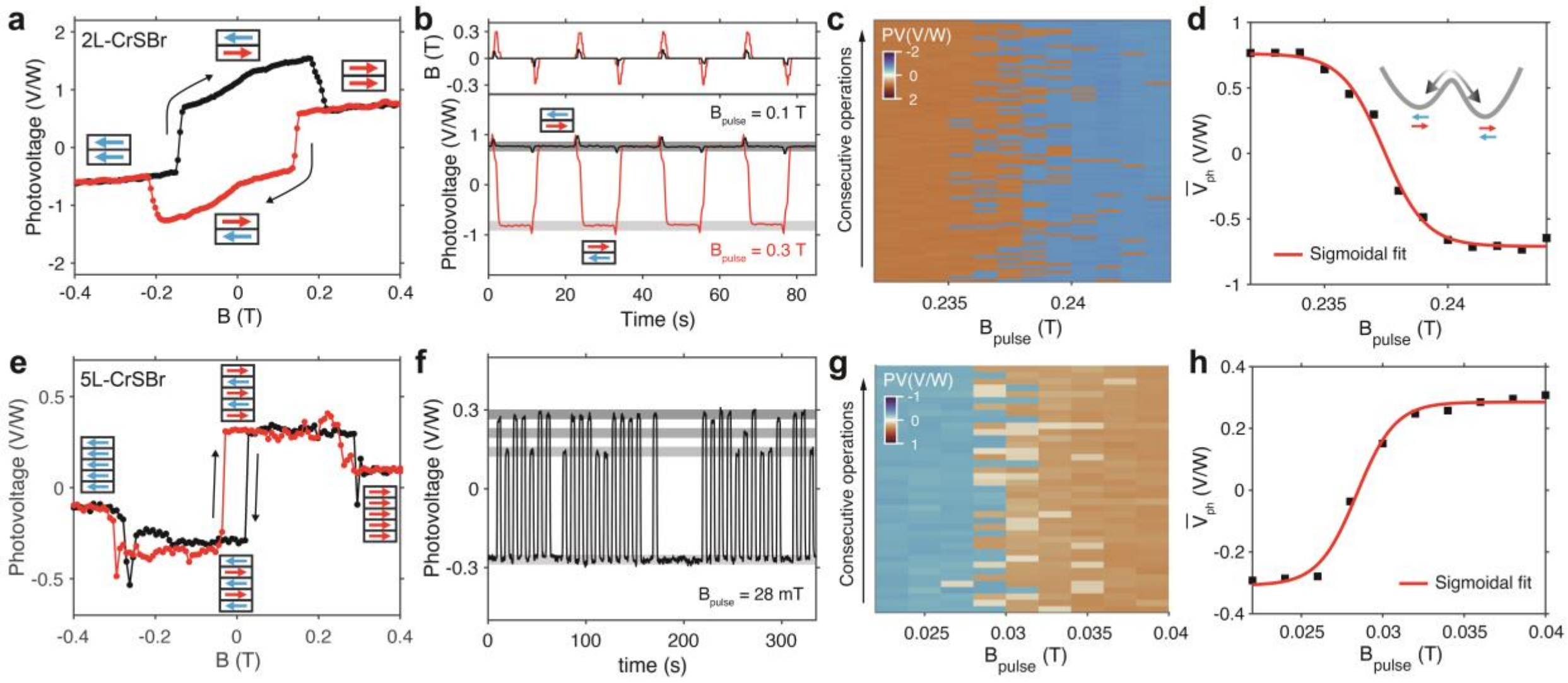

**Fig. 4. Quantum-metric-driven reconfigurable, nonvolatile and probabilistic photodetection a**, Photovoltage as function of magnetic field in a 2L-CrSBr device with excitation photon energy 1.337eV. **b**, Manipulation of photoresponsivity states with magnetic field pulse. The magnetic field pulse and the corresponding time evolution of photovoltage are shown in the upper and lower panel. **c**, Measurement of the photovoltage after magnetic field pulse for series of consecutive operations in the 2L-CrSBr device. A reversed pulse ($B_{pulse} = -0.35\,T$ ) is applied before each operation to initialize the device to the positive photoresponsivity state. **d**, Statistic average photovoltage $\bar{V}_{ph}$ as function of magnetic field pulse strength $B_{pulse}$. Inset: illustration for tuning of probabilistic switching. **e**, Photovoltage as function of magnetic field in a 5L-CrSBr device with excitation photon energy 1.359 eV. **f**, Demonstration of stochastic switching and multiple photoresponsivity states in the 5L-CrSBr device. **g**, Measurement of the photovoltage after magnetic field pulse for series of consecutive operations in the 5L-CrSBr device. An initialization pulse of $B_{pulse} = -0.15\,T$ is applied before each operation. **h**, $\bar{V}_{ph}$ as function of $B_{pulse}$ in 5L sample. All the measurements are conducted at 10 K and the spin-state-independent photovoltage component is removed for clearance.

Since $\boldsymbol{v} \cdot g$ can be tuned by magnetic states, it fertilizes the idea of a nonvolatile and reprogrammable magnetic photodetector for in-sensor computing [40]. We firstly examine the idea in a 2L-CrSBr device, in which the macroscopic quantum-metric response is symmetry-allowed. As shown in Fig. 4a, the 2L-CrSBr device exhibits reproducible switching between positive and negative photoresponsivity states by sweeping of magnetic fields. Fig. 4b further shows the on-demand setting of photoresponsivity states with magnetic pulses ($B_{pulse} = 0.3\ T$). The nonvolatile photoresponsivity states are stable over time and show strong noise tolerance as illustrated with weaker magnetic pulse ($B_{pulse} = 0.1\ T$). For intermediate pulse strength $B_{pulse} \sim 0.238\ T$, a probabilistic photodetection is found that the photovoltage after magnetic pulse for repeated consecutive operations appears randomly for the positive and negative photoresponsivity states in Fig. 4c. Correspondingly, the averaged photovoltage $\bar{V}_{ph}$ can be well described by a sigmoidal function $\bar{V}_{ph} = A\left[1 - e^{(B_{pulse} - B_0)/w}\right] / \left[1 + e^{(B_{pulse} - B_0)/w}\right]$ of $B_{pulse}$, showing a continuously tunable probabilistic photoresponse (Fig. 4d). These results prove the concept for a nonvolatile, reprogrammable and probabilistic quantum-metric-driven photodetection with the benchmarking performance in the symmetry-allowed 2L-CrSBr device.

We then turn to a 5L-CrSBr device, in which the macroscopic quantum-metric response is symmetry-forbidden and the hidden quantum metric response is dominant. Unlike the relatively simple bistable switching in the 2L sample, the 5L sample exhibits richer multilevel photoresponsivity and more complex switching kinetics (Fig. 4e,f,g). Despite this more complicated switching landscape, $\bar{V}_{ph}$ can be continuously tuned with $B_{pulse}$ and the $\bar{V}_{ph}$-$B_{pulse}$ relation can be likewise described by a sigmoidal function (Fig. 4h). This indicates that, at the level of the statistic averaging, the centrosymmetric 5L-CrSBr device realizes the same class of probabilistic photoresponse as the 2L-CrSBr benchmark. It therefore suggests that such quantum-metric-driven innovated photodetector can be realized not only in the conventional quantum-metric regime, but also in centrosymmetric magnetic phases through the hidden-response channel.

It needs to be noted that the magnetic-field controlling knob can be replaced with field-free spin-orbit torques switching of the Néel vector [41-43], which promises higher switching speed and endurance than charge-based devices. Such nonvolatile, reprogrammable and probabilistic photodetection is of broad interest for emerging sensing and computing paradigms, particularly those involving probabilistic sensing, in-sensor computing and neuromorphic vision [44,45].

## Conclusions

To summarize, our work reveals that conventional symmetry perspective of the quantum geometric properties overlooks a wealth of hidden quantum-metric-driven effects in large class of materials with inversion symmetry. We showcase the extraction of the hidden quantum metric response using a non-equilibrium, non-local photovoltaic scheme. These results not only expand the fundamental landscape for quantum geometry physics, but also demonstrate the potential for encoding the quantum geometric information into spintronic and optoelectronic devices for neuromorphic in-memory sensing and computing.

## Methods

### Fabrication of CrSBr device

The CrSBr crystals were mechanically exfoliated onto a 285-nm $SiO_2$/Si substrate. The yield CrSBr flakes of different layer numbers was identified using optical microscopy. The layer number is further confirmed by atomic force microscopy. Cr/Au (5/25 nm) electrodes were fabricated in advance on the $SiO_2$/Si substrate via standard UV photolithography and magnetron sputtering. The exfoliated CrSBr flakes were then picked up, aligned, and transferred onto the pre-patterned electrodes using a polymer-based dry transfer method [46]. The exfoliation and fabrication of CrSBr devices were conducted in a glove box filled with an inert atmosphere. The as-fabricated CrSBr devices were thermally treated for 10 minutes at 120°C inside the glove box.

**Magneto-reflectance and photovoltage spectroscopy**

The photovoltage spectroscopy was performed with a home-built setup shown schematically in Extended Data Fig. 1. The samples were placed in a closed-cycle cryostat with the external magnetic field applied in-plane. A supercontinuum ns-laser with a repetition rate of 10 MHz is used for optical excitation. The white-light laser is monochromated and collimated. The resulting single-wavelength laser light is focused onto the sample using a 20× non-magnetic objective lens. The laser spot size is estimated to be ~ 2 μm. The excitation polarization direction is changed by the rotation of a wide-band half-wave plate. The position of the laser spot is controlled by a galvo mirror which allows us to perform 2D imaging of the photovoltage signal. The reflected laser light from the sample surface is collected in the backscattering geometry through the same objective lens and detected simultaneously with the photovoltage measurement using a Si diode.

**First-principles calculation**

The first-principles calculations based on density functional theory (DFT) are performed using VASP software [47,48]. The projector-augmented wave (PAW) method and the generalized-gradient approximation (GGA) of exchange-correlation energy functional in the Perdew-Burke-Ernzerhof (PBE) functional is employed; we further adapted the DFT-D3 functional to account for weak van der Waals (vdW) interlayer interactions. The spin-orbit coupling (SOC) effects are considered. $U_{eff}$= 3.0 eV was set for Cr atoms to account for strong electronic correlations.

The DFT Bloch wave functions are iteratively transformed into maximally localized Wannier functions by the Wannier90 package [49], then the effective tight-binding (TB) Hamiltonian is obtained. Cr: $d$, S: $p$, Br: $p$ orbitals are used to construct the Wannier functions. In order to calculate BPVE for the thick layer with an acceptable amount of computation, we construct a slab system [50] which is periodic along $\boldsymbol{a}$ and $\boldsymbol{b}$ directions. We label the layer index along $c$-direction as $i$. As a consequence, the Hamiltonian of a slab system with $n_s$ layers can be written in the layer index matrix form

$$H_{mn}^{\text{slab}}\left(\boldsymbol{k}_{\parallel}\right)=\begin{pmatrix} H_{mn,11}\left(\boldsymbol{k}_{\parallel}\right) & H_{mn,12}\left(\boldsymbol{k}_{\parallel}\right) & \cdots & H_{mn,1n_s}\left(\boldsymbol{k}_{\parallel}\right) \\ H_{mn,21}\left(\boldsymbol{k}_{\parallel}\right) & H_{mn,22}\left(\boldsymbol{k}_{\parallel}\right) & \cdots & H_{mn,2n_s}\left(\boldsymbol{k}_{\parallel}\right) \\ \vdots & \vdots & \ddots & \vdots \\ H_{mn,n_s1}\left(\boldsymbol{k}_{\parallel}\right) & H_{mn,n_s2}\left(\boldsymbol{k}_{\parallel}\right) & \cdots & H_{mn,n_sn_s}\left(\boldsymbol{k}_{\parallel}\right) \end{pmatrix}, \qquad \text{S(1)}$$

where the diagonal elements of the Hamiltonian are the intra-plane ones, and the off-diagonal elements are the inter-plane ones, they are all obtained from the bulk Wannier90 Hamiltonian. The element in Eq. S(1) can be read explicitly as

$$H_{mn,ij}\left(\boldsymbol{k}_{\parallel}\right)=\sum_{\boldsymbol{R}=\{\boldsymbol{R}_1,\boldsymbol{R}_2,(i-j)\boldsymbol{R}_3\}} e^{i\boldsymbol{k}_{\parallel}\cdot\boldsymbol{R}}\, H_{mn}(\boldsymbol{R}). \qquad \text{S(2)}$$

Finally, the slab Hamiltonian can be obtained straightforwardly to diagonalize Eq. S(1).

The BPVE optical current can be described as

$$J^a=\sigma_{bc}^a(0;\omega,-\omega)E_b(\omega)E_c(-\omega), \qquad \text{S(3)}$$

where $E_b$ is the b component of the electric field $\boldsymbol{E}$ of light, and the Einstein summation convention is adopted here.

The linear polarized light irradiated on magnetic materials can induce the BPVE, which arises from the injection current mechanism and is time-reversal odd. The BPVE current coefficients can be calculated as follows [51,52]

$$\sigma_{bc}^a(0,\omega,-\omega)=-\frac{\pi e^3\tau}{4\hbar^2}\int\frac{d^3\boldsymbol{k}}{(2\pi)^3}\sum_{n,m} f_{nm}\Delta_{nm}^a\{r_{mn}^c,r_{nm}^b\}\delta(\omega_{mn}-\omega), \qquad \text{S(4)}$$

where $\Delta_{nm}^{a} = v_{nn}^{a} - v_{mm}^{a}$ represents the band velocity difference, $\{r_{mn}^{c}, r_{nm}^{b}\} = (r_{mn}^{c})^{*} r_{nm}^{b} + (r_{nm}^{b})^{*} r_{mn}^{c}$ is the quantum metric, $\delta(\omega_{mn} - \omega)$ is the delta function indicating that optical transitions occur when the photon energy $\omega$ matches the band energy difference $\omega_{mn}$. The term $\sum_{n,m} f_{nm} \Delta_{nm}^{a} \{r_{mn}^{c}, r_{nm}^{b}\} \delta(\omega_{mn} - \omega)$ corresponds to the interband quantum metric dipole that arises at the optical transition energy, which is denoted as $v_a g_{bc}$ in the following discussion for brevity.

In order to evaluate the contributions of each layer to the BPVE coefficients, we introduce a projection operator $P_i = \sum_{n \in \{L_i\}} |\psi_j\rangle \langle \psi_j|$ where $|\psi_j\rangle$ represent Wannier functions centered at the atoms which belong to the $i$-th layer. The layer projection operation matrix under the Wannier basis can be written as

$$P_i = \begin{bmatrix} \delta_{1i} & 0 & 0 & 0 & 0 \\ 0 & \delta_{2i} & 0 & 0 & 0 \\ 0 & 0 & \delta_{3i} & 0 & 0 \\ 0 & 0 & 0 & \ddots & \vdots \\ 0 & 0 & 0 & \cdots & \delta_{ni} \end{bmatrix}, \qquad \text{S(5)}$$

$\delta_{ni} = 1$ only if the $n$-th Wannier function belongs to the $i$-th layer. $P_1 + P_2 + \cdots + P_n = I_{0,n\times n}$. Then we multiply the velocity operator $\hat{v}_a$ by the projector, namely $\hat{v}_a \rightarrow (P_i \hat{v}_a + \hat{v}_a P_i)/2$ [53,54]. The subscript "$a$" means the direction of optical response in Eq. S(4). Through this projection method, one could map the photocurrent onto each vdW layer.

**Symmetry Analysis**

The magnetic point group of even-layer CrSBr is m'mm, which has the $PT$ symmetry and allows the $T$-odd function nonlinear effect. According to the isomorphic group method [55], the BPVE tensor under the magnetic point group m'mm is equivalent 2mm (the rotation operation 2 is along the $a$-axis), i.e.

$$[\sigma]_{3\times6} = \begin{pmatrix} \sigma_{11} & \sigma_{12} & \sigma_{13} & 0 & 0 & 0 \\ 0 & 0 & 0 & 0 & 0 & \sigma_{26} \\ 0 & 0 & 0 & 0 & \sigma_{35} & 0 \end{pmatrix} \quad \text{S(6)}$$

where the Voigt notation is adopted here. The three independent BPVE coefficients for even-layer CrSBr are $\sigma_{xx}^{x}$, $\sigma_{yy}^{x}$ and $\sigma_{xy}^{y} = \sigma_{yx}^{y}$.

For odd-layer CrSBr, its magnetic point group is m'mm', which has inversion symmetry. The resulting global BPVE is forbidden by symmetry. Nevertheless, it has the hidden BPVE at the top and bottom surface layers because of lower local symmetry. As an example, the projected BPVE coefficient on the surface monolayer is shown in Extended Data Fig. 2. It also reveals the dominant contribution of $\sigma_{yy}^{x}$ over $\sigma_{yy}^{x}$ and $\sigma_{xy}^{y}$ due to the quasi-one-dimensional character of the band-edge electronic structure.

**Evaluation of the Au/CrSBr Schottky contact**

To evaluate the contact between CrSBr and the Au electrode, we calculated the work functions of both materials separately. The Coulomb potential of Au and CrSBr was computed using a slab model, from which the work function can be obtained from the vacuum level. As shown in Extended Data Fig. 5(a), the calculated work function of Au is 5.15 eV, which is in good agreement with typical reported experimental values. Similarly, the work functions of CrSBr in the AFM and FM states are 5.98 eV and 5.94 eV, respectively. These results confirm that a stronger Schottky barrier is formed between Au electrode and AFM CrSBr due to larger work function difference as comparing to FM phase. This is consistent with the experiments observation of strong spin-state-independent photovoltage component in the close vicinity of the CrSBr/Au contact for AFM phase as shown in Extended Data Fig. 4. The spin-state-independent photovoltage is driven by the built-in electrical fields at the Schottky junction, which has opposite sign for left and right contact. The formation of Schottky junction is also confirmed by the temperature dependent IV characterization of the devices shown in Extended Data Fig. 6. A weak but finite Schottky barrier is obtained by fitting of the Arrhenius plot of the IV characteristics with thermionic emission model [56].

**Data Availability**

All data generated or analysed during this study are included in this published article.

**Acknowledgments** This work was financially supported by the National Key Research and Development Program of China (No. 2022YFA1405100, 2022YFA1204003, 2024YFB3614100), the National Natural Science Foundation of China (No. 12241405, 12427805, 12374077, 12174384, 12174378, 12274406, 12274411, 12474100, 12204009), the

Basic Research Program of the Chinese Academy of Sciences Based on Major Scientific Infrastructures (Grant No. JZHKYPT-2021-08) and the CAS Project for Young Scientists in Basic Research (No. YSBR-120). The calculations were performed at Hefei Advanced Computing Center.

**Author contributions** Y.H., D.-F.S. and K.W. conceived and coordinated the project. Y.T. and Q.H. fabricated the device and performed characterization under the supervision of Y.H., L. Z. and K.W., Y.T., Q.H. and Y.H. performed photovoltaic spectroscopy experiments and analyzed the data. R.-C.X., H.Z. and D.-F.S. carried out the theoretical analyses and calculations. Y.T., Q.H., Y.H., D.-F.S. and K.W. wrote the manuscript, with contributions from all other coauthors.

**Competing interests** The authors declare no competing interests.

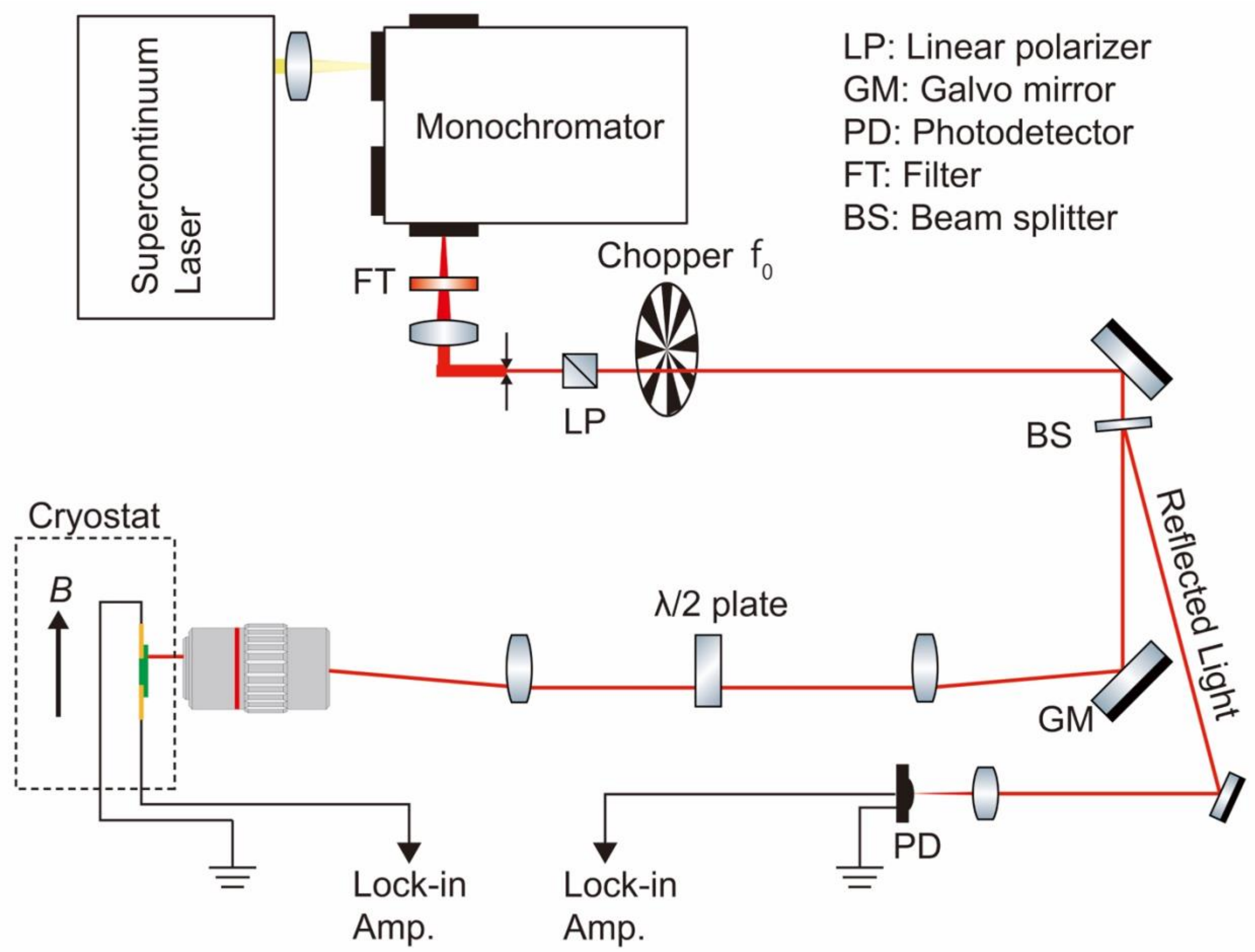

**Extended Data Fig. 1. Optical set-up for magneto-reflectance and photovoltage spectroscopy measurement.**

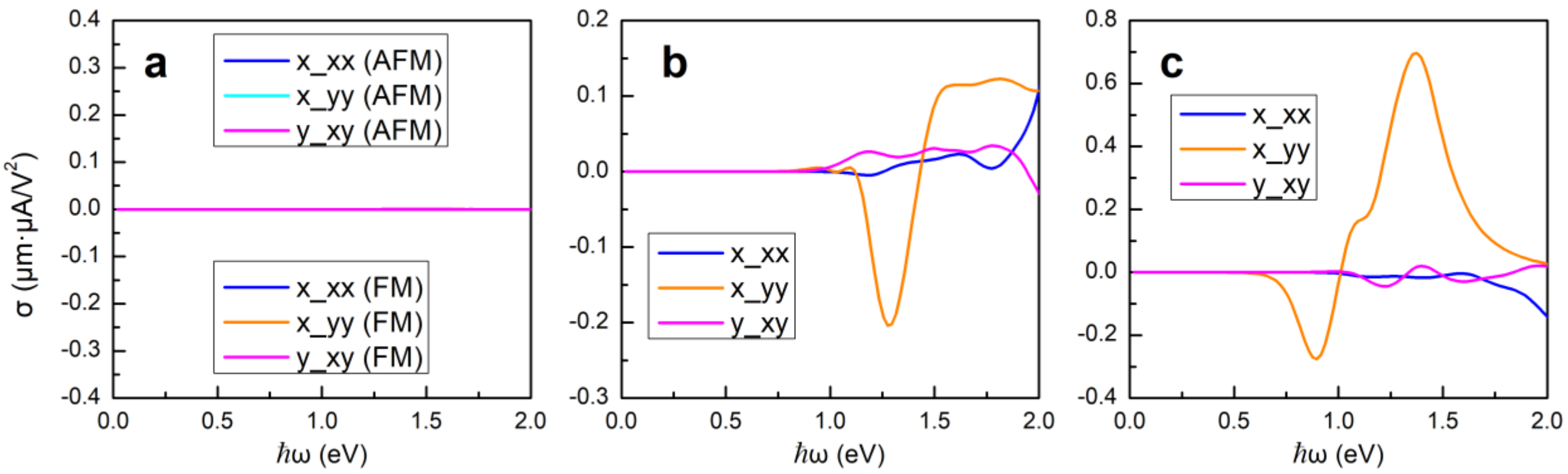

**Extended Data Fig. 2. The total and projected BVPE coefficients for 5L-CrSBr slab**

**a**, The BVPE coefficients $\sigma_{xx}^{x}$, $\sigma_{yy}^{x}$ and $\sigma_{xy}^{y}$ for 5L AFM and FM state. **b**,**c**, The projected BPVE coefficients on the surface monolayer of 5L-CrSBr for the AFM and FM states, respectively.

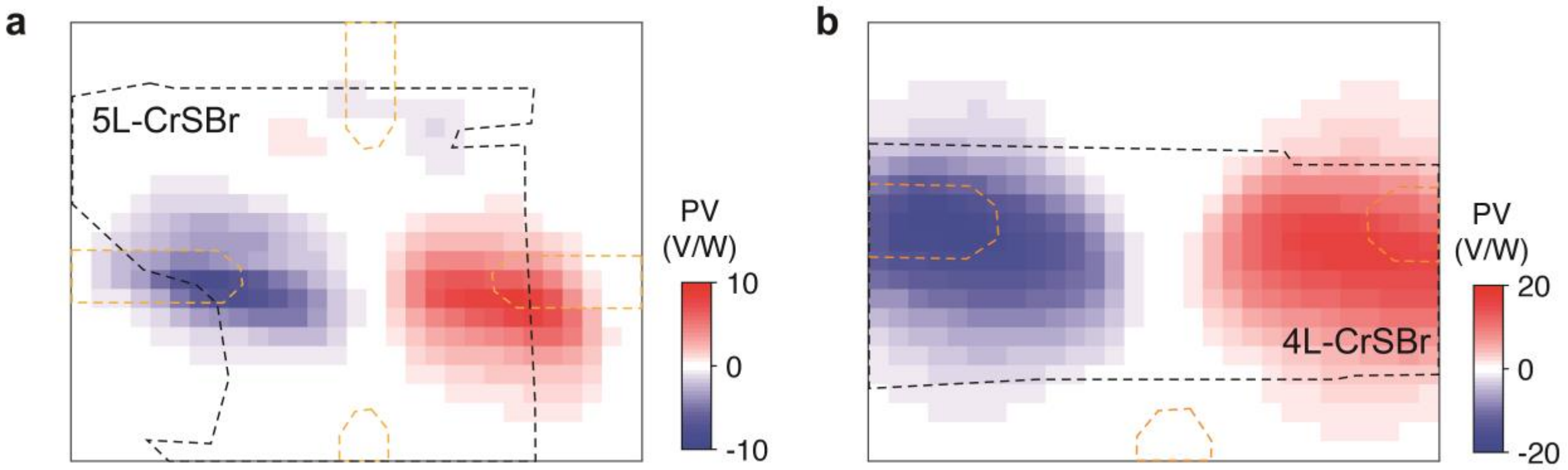

**Extended Data Fig. 3. spatial distribution of the spin-state-independent photovoltage component**

**a** and **b** are the spatial mapping of the spin-state-independent photovoltage in 5L- and 4L-CrSBr device at B = 0 under resonant excitation of X1. The spin-state-independent photovoltage is obtained by averaging the photovoltage detected at B = 0 T under forward and backward field sweep. The Au contact and the CrSBr flake are marked with dashed lines. The spacing between Au electrodes in both devices is 20 μm.

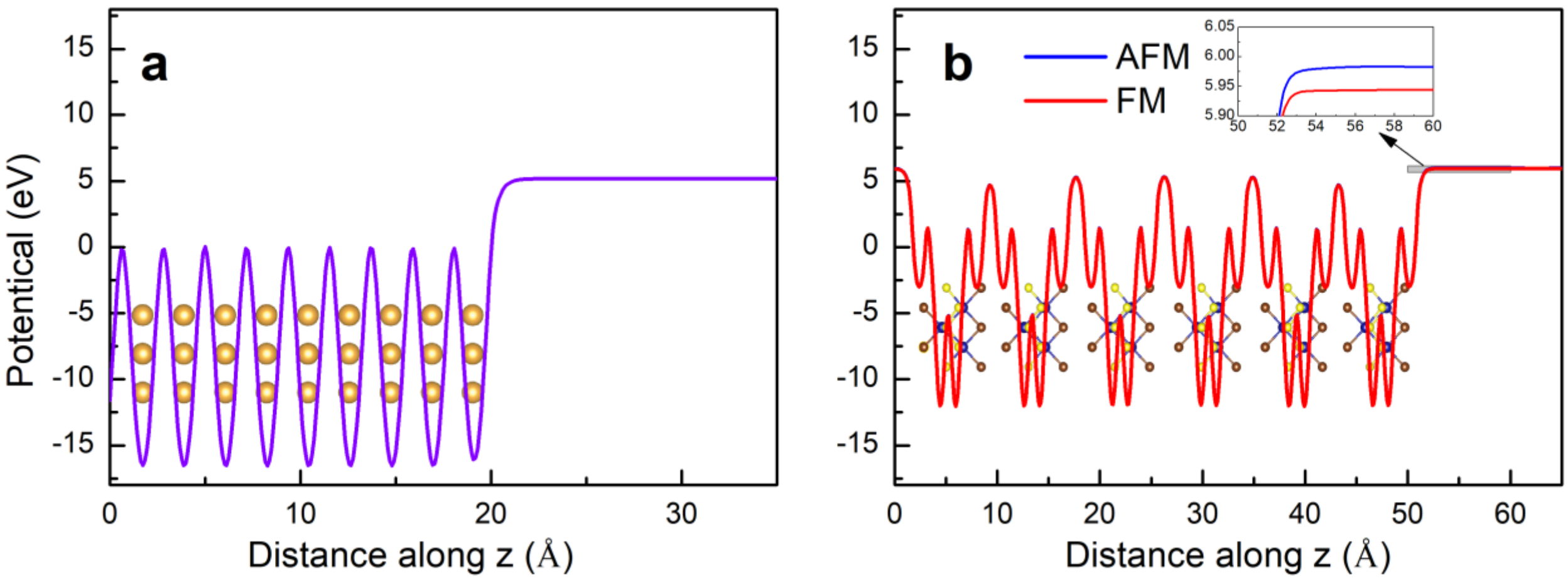

**Extended Data Fig. 4. Evaluation of the Au and CrSBr workfunction**

Calculated Coulomb potential of (**a**) Au and (**b**) CrSBr with AFM and FM states. The Fermi energies are set to zero. The work functions of Au, AFM-CrSBr, and FM-CrSBr are calculated to be 5.15 eV, 5.98 eV, and 5.94 eV, respectively.

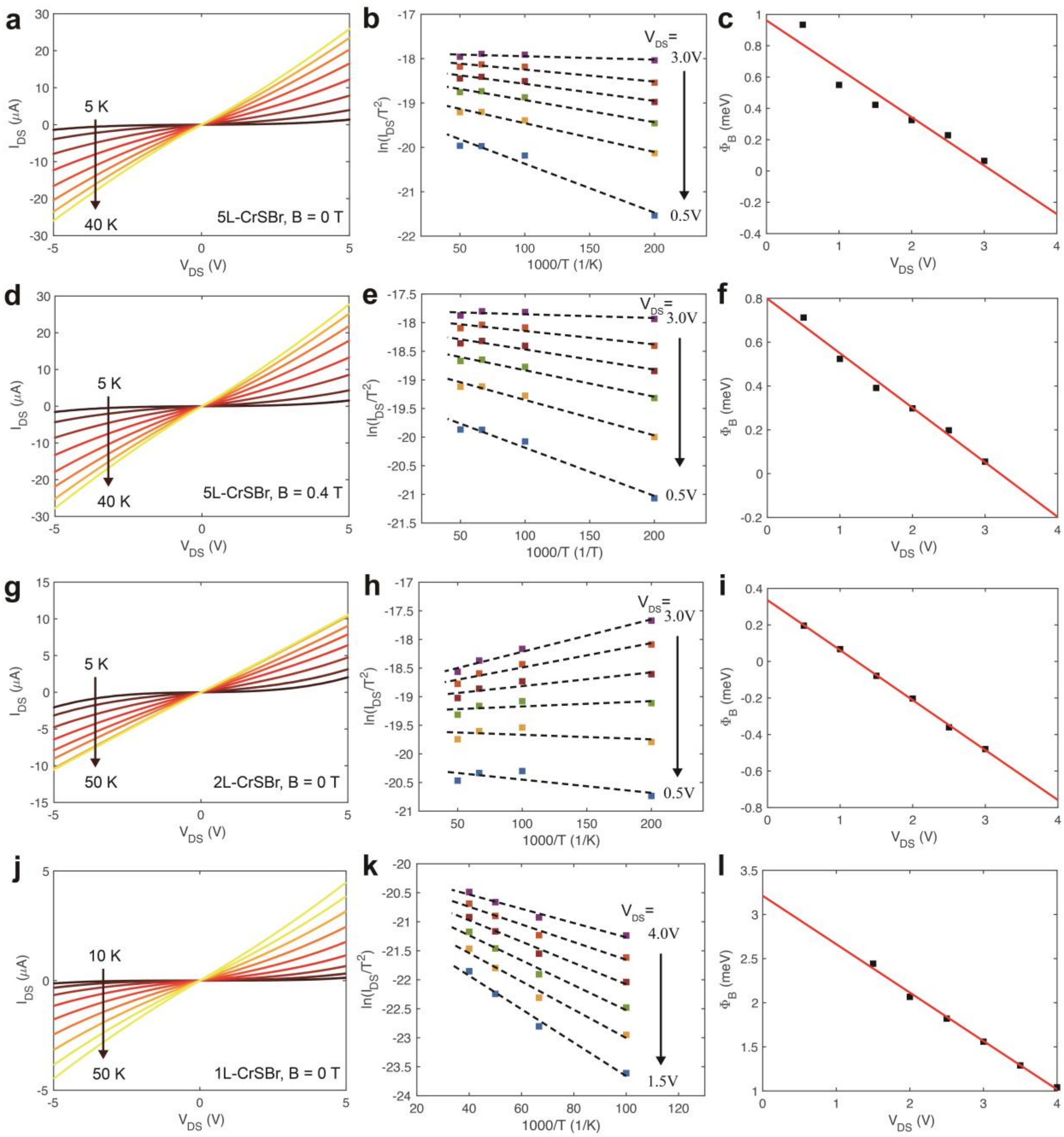

**Extended Data Fig. 5. Temperature dependent IV characterization of the devices.**

**a**, I-V curves of the 5L-CrSBr device measured at various temperatures from 5 K to 40 K in the antiferromagnetic phase at $B$ = 0 T. **b**, Arrhenius plot of the source-drain current $I_{DS}$ for selected bias voltage $V_{DS}$ from **a**. **c**, Estimation of the Schottky barrier height $\Phi_B$ by fitting the slope from **b**. **d**, Temperature dependent I-V curves of the 5L-CrSBr device for the ferromagnetic phase by setting $B$ = 0.4 T. **e**, Arrhenius plot of $I_{DS}$ for selected $V_{DS}$ from **d**. **f**, Estimation of $\Phi_B$ by fitting the slope from **e**. **g** and **j** are the temperature dependent I-V curves of the 2L- and 1L-CrSBr device at $B = 0$ T. **h** and **k** are the corresponding Arrhenius plot. **i** and **l** are the estimated $\Phi_B$ respectively.

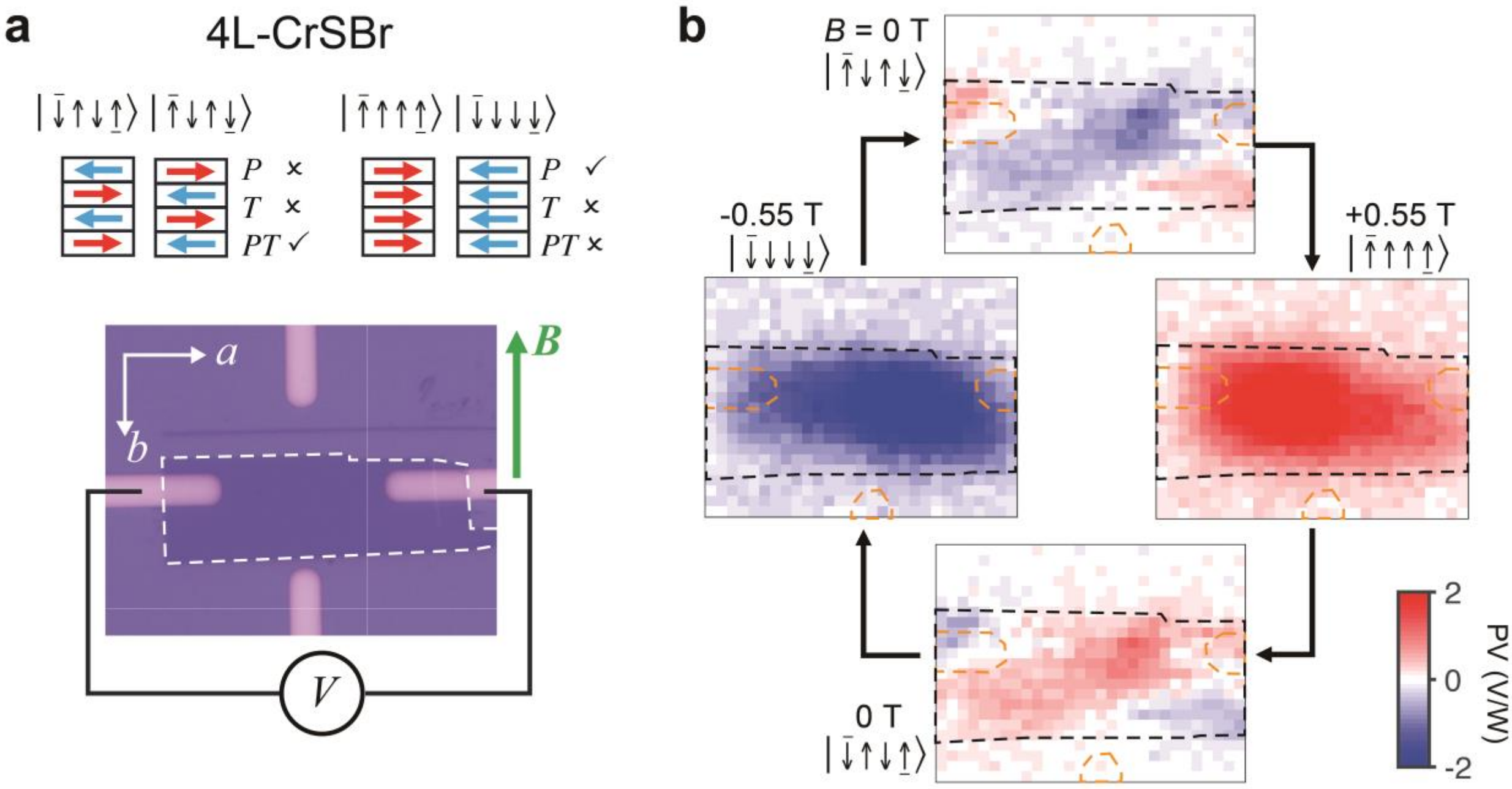

**Extended Data Fig. 6. Spatial Mapping of the magnetic photovoltaic response in 4L-CrSBr device**

**a**, Optical microscopy image of the 4L-CrSBr device. The plausible AFM and FM states are illustrated with the corresponding symmetries. **b**, Spatial mapping of the magnetic photovoltage in 4L-CrSBr under resonant excitation of $X_1$ and its evolution with magnetic fields. The Au contact and the CrSBr flake are marked with dashed lines in the photovoltage 2D mapping. The spacing between Au electrodes in both devices is 20 μm.

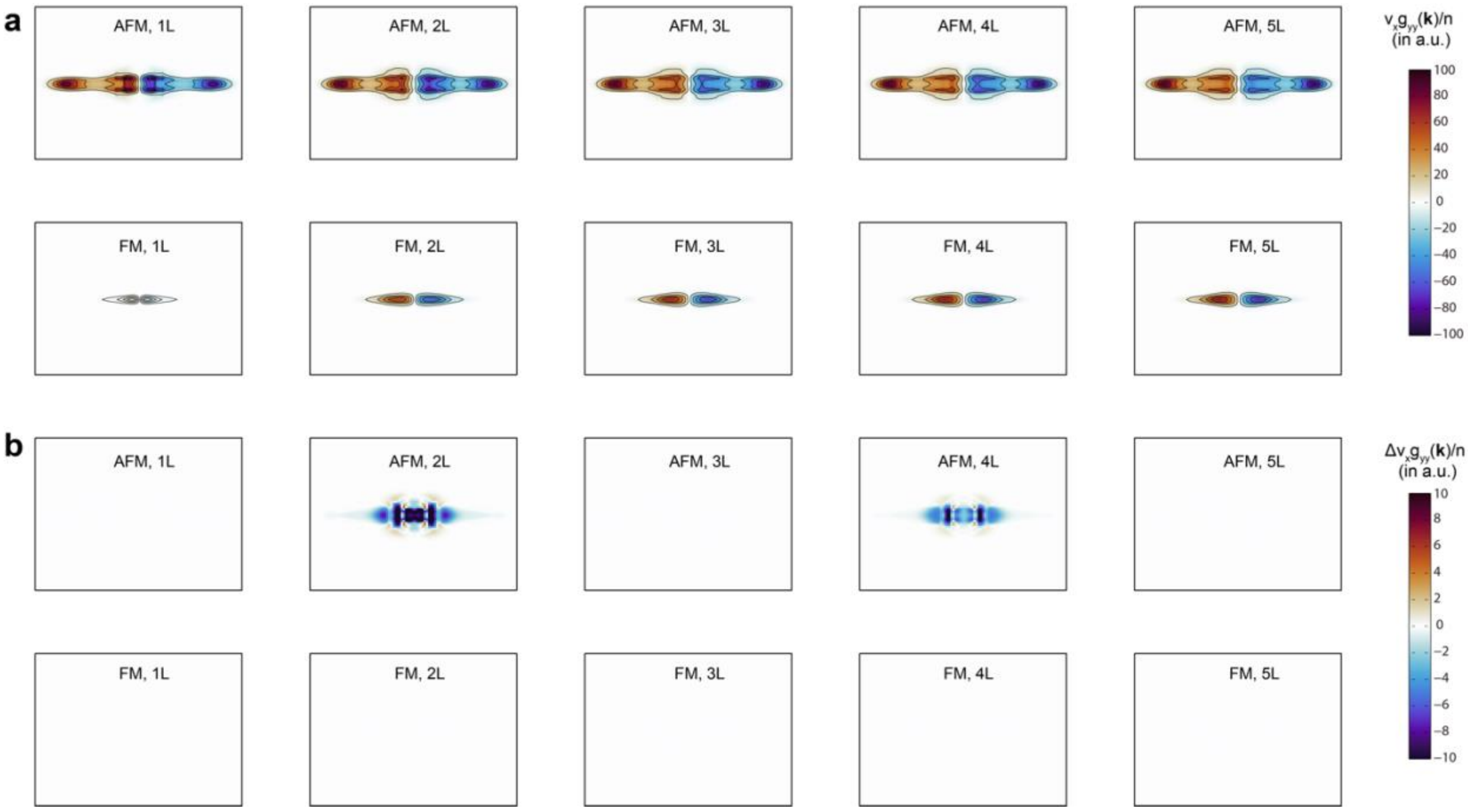

**Extended Data Fig. 7. Calculation of the global quantum metric dipole in CrSBr with varying thickness**

**a**, The calculated momentum-space distribution of the total quantum metric $v_x g_{yy}(\boldsymbol{k})/n$ for CrSBr with different magnetic states and layer numbers $\boldsymbol{n}$. **b**, The calculated momentum-space distribution of the $\Delta v_x g_{yy}(\boldsymbol{k}) = \left[v_x g_{yy}(\boldsymbol{k}) - \mathcal{T}\left(v_x g_{yy}(\boldsymbol{k})\right)\right]/n$ for CrSBr with different magnetic states and layer numbers $\boldsymbol{n}$. The interband transition energies of 1.30 eV and 0.9 eV are used for AFM and FM states calculations, corresponding respectively to the first prominent BPVE peak with lowest transition energies in Extended Data Fig. 2.

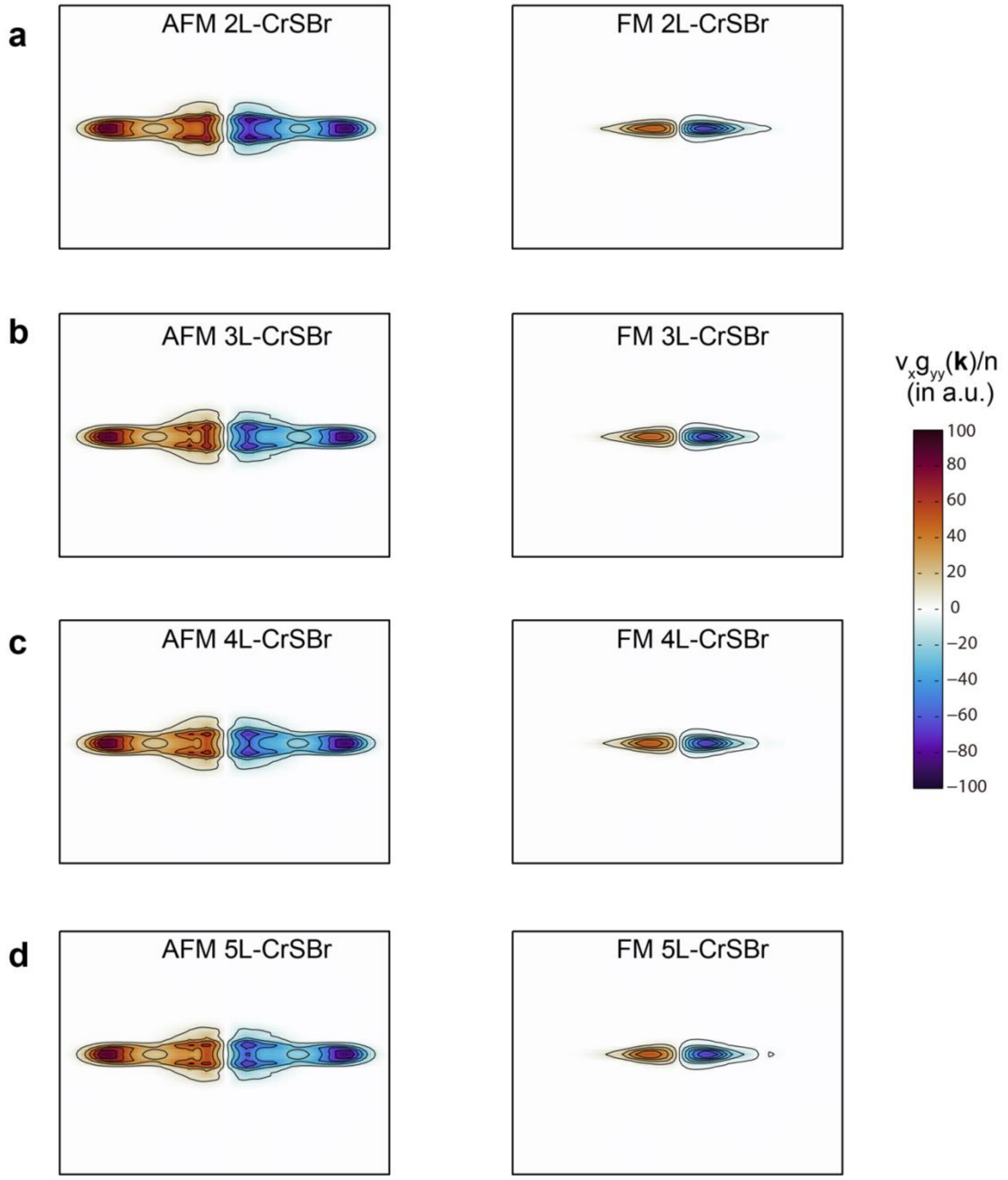

**Extended Data Fig. 8. Calculated projected quantum metric dipole from surface monolayer of 2L-5L CrSBr slab**

**(a-d)** The calculated momentum-space distribution of the projected quantum metric $\boldsymbol{v_x g_{yy}(k)}$ from surface monolayer of a 2L-5L CrSBr in AFM (left) and FM (right) states. The interband transition energies of 1.30 eV and 0.9 eV are used for AFM and FM states calculations, corresponding respectively to the first prominent BPVE peak with lowest transition energies in Extended Data Fig. 2.